\documentclass[twocolumn,aps,superscriptaddress,showpacs,nofootinbib,floatfix]{revtex4}

\usepackage{epsfig,bm,feynmf}

\usepackage{graphics}

\usepackage[normalem]{ulem}  
\usepackage[dvips]{color} 

\renewcommand\sout{\bgroup \color{red} \ULdepth=-.5ex \ULset}

\begin{document}

\title{Effects of initial state fluctuations on jet energy loss}

\author{Hanzhong Zhang}\email{zhanghz@iopp.ccnu.edu.cn}
\affiliation{Institute of Particle Physics and Key Laboratory of Quark $\&$ Lepton Physics, Central China Normal University, Wuhan 430079, China}
\affiliation{Cyclotron Institute, Texas A$\&$M University, College Station, TX 77843-3366, USA}

\author{Taesoo Song}\email{songtsoo@yonsei.ac.kr}
\affiliation{Cyclotron Institute, Texas A$\&$M University, College Station, TX 77843-3366, USA}

\author{Che Ming Ko}\email{ko@comp.tamu.edu}
\affiliation{Cyclotron Institute and Department of Physics and Astronomy, Texas A$\&$M University, College Station, TX 77843-3366, USA}

\date{\today}

\begin{abstract}
The effect of initial state fluctuations on jet energy loss in relativistic heavy-ion collisions is studied in a 2+1 dimension ideal hydrodynamic model. Within the next-to-leading order perturbative QCD description of
hard scatterings, we find that a jet loses slightly more energy in the expanding quark-gluon plasma if the latter is described by the hydrodynamic evolution with fluctuating initial conditions compared to the case with smooth initial conditions. A detailed analysis indicates that this is mainly due to the positive correlation between the fluctuation in the production probability of parton jets from initial nucleon-nucleon hard collisions and the fluctuation in the medium density along the path traversed by the jet. This effect is larger in non-central than in central relativistic heavy ion collisions and also for jet energy loss that has a linear than a quadratic dependence on its path length in the medium.
\end{abstract}

\pacs{12.38.Mh, 24.85.+p; 25.75.-q}

\maketitle

\section{Introduction}

The observation of jet quenching through the suppression of large transverse momentum single hadron, dihadron, and $\gamma$-hadron spectra in relativistic heavy ion collisions \cite{star-suppression,phenix-suppression} is one of the most important evidence for the formation of a strongly coupled quark-gluon plasma (QGP) in these collisions. Jet quenching is a measure of the energy loss of an initial leading jet parton as it traverses through the produced dense matter via multiple scatterings \cite{Wang:1991xy}. Theoretical studies on parton jet energy loss have concentrated on both gluon radiation induced by multiple scattering and elastic collision energy loss. Due to the non-Abelian Landau-Pomeranchuk-Migdal (LPM) interference effect \cite{LPM}, the radiative energy loss shows a quadratic path-length dependence \cite{BDPMS,Zakharov,GLV,GuoW,Wied}, which is in contrast to the linear path-length dependence of the elastic collision energy loss \cite{Mustafa:2004dr,Adil:2006ei,Qin:2007rn}. Also, a cubic path-length dependence of the jet energy loss has been found in the strongly coupled limit of the QCD medium using the AdS/CFT correspondence \cite{Dominguez:2008vd,Marquet:2009eq}.

The study of jet quenching in heavy ion collisions has been carried out in the 1+1 dimension Bjorken hydrodynamics \cite{Zhang:2007ja,Zhang:2009rn, Vitev:2002pf,Kang:2011rt} as well as the 2+1 and 3+1 dimension ideal and viscous hydrodynamics \cite{Chen:2011vt,Qin:2009uh,Renk:2006pk,Renk:2010qx}. In these studies, the initial conditions for the hydrodynamical evolution were taken to be smooth in space. Recently, the effect of initial event-by-event fluctuations on jet quenching has been investigated in the 1+1 Bjorken hydrodynamics \cite{Rodriguez:2010di}. It was found that the strong correlation between the fluctuation in the spatial distribution of initial hard scatterings from which jets are produced and the fluctuation in the density distribution of the initial medium has significant effects on jet quenching. In particular, the jet energy loss is reduced after the inclusion of initial fluctuations.  However, the transverse expansion of the produced hot dense medium has been neglected in this study. In the present paper, we include the transverse expansion in studying the effect of initial fluctuations on jet energy loss by using the 2+1 dimension ideal hydrodynamic model of Refs. \cite{Song:2011xi,Song:2010ix}. For calculating the hadron spectra at large transverse momentum, we use the next-to-leading order (NLO) perturbative QCD. Our results show that including the transverse expansion of the medium slightly enhances the energy loss of jets, contrary to the reduced jet energy loss found in Ref.~\cite{Rodriguez:2010di} without the transverse expansion. We further investigate the effect of initial fluctuations for different path-length dependence of jet energy loss in the medium.

This paper is organized as follows. We first give a brief description of the the 2+1 dimension ideal hydrodynamic model in Sec. \ref{hydro} and the jet quenching models in Sec. \ref{model}. Results from our study are shown in Sec. \ref{results}. We then present some discussions in Sec. \ref{discussions} and finally summarize our study in Sec. \ref{summary}.

\section{2+1 dimension hydrodynamics}\label{hydro}

In the 2+1 dimension ideal hydrodynamics, which assumes the boost invariance along the longitudinal direction, the energy-momentum tensor $T^{\mu\nu}$ and pressure $p$ of a system can be expressed in terms of the proper time $\tau$ and the two transverse coordinates $x$ and $y$ perpendicular to the beam direction \cite{Teaney:2001av,Heinz:2005bw,Song:2011xi,Song:2010ix}. Conservations of energy and momentum then give
\begin{eqnarray}
\partial_\tau (\tau T^{00})+\partial_x (\tau T^{0x})+\partial_y (\tau T^{0y})&=&-p,\nonumber\\
\partial_\tau (\tau T^{0x})+\partial_x (\tau T^{xx})+\partial_y (\tau T^{xy})&=&0,\nonumber\\
\partial_\tau (\tau T^{0y})+\partial_x (\tau T^{xy})+\partial_y (\tau T^{yy})&=&0.
\label{conservations}
\end{eqnarray}
To solve these equations requires information on the initial conditions
of a collision, particularly the initial entropy density, and the equation of state of the produced matter. For the initial entropy density, it is taken as
\begin{eqnarray}
\frac{ds}{d\eta}=C\bigg\{(1-\xi)\frac{n_{\rm part}}{2}+\xi~n_{\rm coll}\bigg\},
\label{initial}
\end{eqnarray}
where $n_{\rm part}$ and $n_{\rm coll}$ are the number densities of participants and binary collisions, respectively.

In heavy ion collisions, the initial conditions vary from event to event as the positions of colliding nucleons are randomly distributed according to the density distributions of the colliding nuclei. Two nucleons are considered as participants and a binary collision takes place at their middle point if the transverse distance between a nucleon from one nucleus and a nucleon from the other nucleus is shorter than $\sqrt{\sigma_{\rm in}/\pi}$, where $\sigma_{\rm in}=42~{\rm mb}$ is the nucleon-nucleon inelastic cross section at RHIC energies. A smearing parameter $\sigma$ is then introduced in evaluating the number densities of participants and binary collisions, i.e.,
\begin{eqnarray}
n_{{\rm part}({\rm coll})}({\bf r})&=&\frac{1}{2\pi\sigma^2\tau_0}\sum_{i=1}^{N_{{\rm part}({\rm coll})}}\exp\bigg(-\frac{|{\bf r}_i-{\bf r}|^2}{2\sigma^2}\bigg),
\label{numbers2}
\end{eqnarray}
where ${\bf r}_i$ is the transversal position of a participant (binary collision). Here we use the same smearing parameter $\sigma$ for both the participant and binary collision number densities. In the present study, we consider the two cases of $\sigma$ = 0.4 fm and 0.8 fm. Also, we choose the initial thermalization time $\tau_0$ = 0.6 fm/c for starting the hydrodynamical evolution.

For the equations of state, we use the quasi-particle model based on the lattice QCD data for the QGP and the resonance gas model for the hadron gas as in Refs.~\cite{Levai:1997yx,Song:2011xi,Song:2010ix}. This model thus assumes the presence of a first-order phase transition and the critical temperature $T_c$ is 170 MeV. We solve the hydrodynamic equations Eq. (\ref{conservations}) numerically by using the HLLE algorithm \cite{Schneider:1993gd,Rischke:1995ir,Rishke:1998}.

The parameters $C$ and $\xi$ in Eq.~(\ref{initial}) are determined from fitting the centrality dependence of the final charged-particle multiplicity \cite{Back:2004je}. Using the Cooper-Frye freeze-out formula and assuming that the multiplicity does not change after chemical freeze-out at temperature $T= 160$ MeV \cite{Song:2011xi}, we obtain $C=19.3$ and $\xi=0.11$.

In studies with smooth initial conditions, both the participant number and the binary collision number densities are obtained from the thickness functions of the colliding nuclei evaluated from their density distributions. In the present study, they are obtained by averaging over a large number of initial fluctuating events. Because of the smearing parameter introduced in generating the initial conditions for hydrodynamical evolutions, the resulting smooth participant number and the binary collision number densities have a larger spread in space than that obtained from the nuclear thickness functions.

\section{Jet queching models}\label{model}

For a jet of energy $E$ produced at the position ${\bf r}$ from a hard nucleon-nucleon collision and moving along an azimuthal angle $\phi$ in the transverse plane of a nucleus-nucleus collision, its total energy loss can be expressed as
\begin{eqnarray}
\Delta E = \int d\tau f(E,\phi,{\bf r},\tau) \rho({\bf r},\phi,\tau), \label{eqn:enloss_path_integ}
\end{eqnarray}
where $\rho({\bf r},\phi,\tau)$ is the local parton density at time $\tau$ along the jet path, and the function $f(E,\phi,{\bf r},\tau)$ is the jet energy loss per unit time through a unit density of medium.

Averaging over the creation positions and moving directions of the jet in the transverse plane gives
\begin{eqnarray}
\langle \Delta E\rangle = \frac{1}{2\pi}\int d\phi d^2{\bf r} d\tau n({\bf r})f(E,\phi,{\bf r},\tau) \rho({\bf r},\phi,\tau), \label{eqn:enloss_sm}
\end{eqnarray}
where $n({\bf r})=n_{\rm coll}({\bf r})/N_{\rm bin}$, with $n_{\rm coll}({\bf r})$ and $N_{\rm coll}$ denoting, respectively, the number density of binary collisions at {\bf r} and the total number of binary collisions, is the probability density for jet production at {\bf r}. The average jet energy loss rate along the jet path is then
\begin{eqnarray}
\frac{d \langle\Delta E\rangle}{d\tau} = \frac{1}{2\pi}\int d\phi d^2{\bf r} n({\bf r}) f(E,\phi,{\bf r},\tau) \rho({\bf r},\phi,\tau). \label{eqn:enloss_time}
\end{eqnarray}

According to recent theoretical studies \cite{Zhang:2007ja,Zhang:2009rn,gvw,ww,Salgado:2002cd}, the total quark energy loss in a finite and expanding medium is approximately given by
\begin{eqnarray}
\Delta E =\left\langle \frac{d E}{d L} \right\rangle\int_{\tau_0}^{\infty} d\tau\left ( \frac{\tau-\tau_0}{\tau_0} \right )^{\alpha}
\frac{\rho(\tau,{\bf r})}{\rho(\tau_0,0)}\frac{p^\mu u_\mu}{p_0}.
\label{eq:de-twist}
\end{eqnarray}
In the above, $\alpha$ is a parameter with possible values 0, 1 and 2,
corresponding, respectively, to linear, quadratic, and cubic path-length dependence for the jet energy loss; and $p^\mu$ and $u^\mu$ are, respectively, the four momentum of the jet and the four flow velocity of the local medium. The average energy loss per unit length $\langle dE/dL\rangle$ has the following parametrization~\cite{ww}:
\begin{eqnarray}
\left \langle\frac{dE}{dL} \right \rangle=\epsilon_0
(E/\mu_0-1.6)^{1.2}/(7.5+E/\mu_0),
\label{eq:loss}
\end{eqnarray}
where $\epsilon_0$ is the energy loss parameter with a value that is $9/4$ times larger for a gluon than for a quark, and $\mu_0$ is the Debye mass. Their values are determined from fitting the experimental data for the nuclear modification factor in the most central $A+A$ collisions using the smooth initial conditions. This leads to the following jet energy loss rate through a unit density of medium:
\begin{eqnarray}\label{energyloss}
f(E,\phi,{\bf r},\tau)=\left\langle\frac{dE}{dL}\right\rangle\left(\frac{\tau-\tau_0}{\tau_0}\right)^\alpha \frac{1}{\rho(\tau_0,0)}\frac{p^\mu u_\mu}{p_0}
\end{eqnarray}

\section{Results}\label{results}

In this Section, we present results for the nuclear modification factor of jets in both central and mid-central Au+Au collisions at $\sqrt{s_{NN}}=200$ GeV. For a given event $i$, the nuclear modification factor is defined as
\begin{eqnarray}
R_{AA}^{i} = \frac{d\sigma_{AA}^i/dp_T^2dy}{N_{\rm bin}^i d\sigma_{NN}/dp_T^2dy}, \label{eqn:modifactoer_fl}
\end{eqnarray}
where the hadron spectra in heavy-ion collisions $d\sigma_{AA}^i/dp_T^2dy$ are calculated from the NLO pQCD with modified fragmentation functions due to jet quenching \cite{Zhang:2007ja,Zhang:2009rn}. Specifically, the cross sections for the hard scattering are calculated using the CTEQ6M parameterizations \cite{distribution} for the parton distributions in a nucleon and including both $2\rightarrow 3$ tree level contributions and 1-loop virtual corrections to $2\rightarrow2$ tree processes \cite{Owens}. Furthermore, the AKK08 parameterizations \cite{akk08} are used for the parton fragmentation into hadrons.

The nuclear modification factor for the case of the fluctuating initial conditions is then given by the average of Eq.~(\ref{eqn:modifactoer_fl}) over all events. For the case of smooth initial conditions, the nuclear modification factor is obtained, on the other hand, from the ratio of corresponding charged hadron spectrum to that of the proton-proton collision multiplied by the average number $N_{\rm coll}$ of nucleon-nucleon collisions.

\begin{figure}[h]
\includegraphics[width=0.9\linewidth]{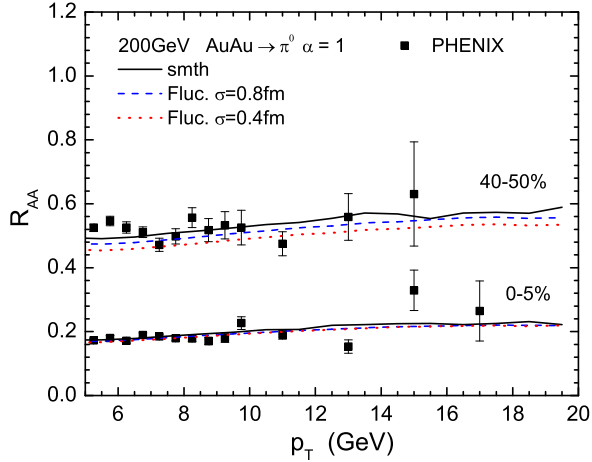}
\caption{\label{fig:Raa-rhic-05-4050} (Color online). Nuclear modification factors with and without initial fluctuations in 0-5\% and 40-50\% centralities of Au+Au collisions at $\sqrt{s}=$ 200 GeV, respectively. The experimental data are taken from \cite{Adare:2008qa}.}
\end{figure}

Fig.~\ref{fig:Raa-rhic-05-4050} shows the nuclear modification factors of high-$p_T$ particles in 0-5\% and 40-50\% centralities of Au+Au collisions at $\sqrt{s}=$ 200 GeV with and without initial fluctuations. It is seen that including initial fluctuations leads to a smaller $R_{AA}$, and the effect is stronger for larger initial fluctuations (corresponding to smaller $\sigma$) and in noncentral than in
central collisions.

\section{discussions}\label{discussions}

The smaller $R_{AA}$ in the case of fluctuating initial conditions obtained in the present study is opposite to the result reported in a previous study based on the 1+1 boost invariant hydrodynamics \cite{Rodriguez:2010di}, where a larger $R_{AA}$ was obtained when initial fluctuations were included. To understand this difference, we define the jet energy loss difference $\delta \langle\Delta E\rangle \equiv \langle\Delta E\rangle^{\rm fluc} - \langle\Delta E\rangle^{\rm smth}$ between the average energy loss calculated with fluctuating initial conditions and that with smooth initial conditions. From Eq. (\ref{eqn:enloss_time}), we then obtain the following rate for this difference along the jet path:
\begin{eqnarray}\label{deviation}
\frac{d[\delta\langle\Delta E\rangle]}{d\tau}
=\frac{1}{2\pi}\int d\phi d^2 {\bf r} f(E,\phi,{\bf r},\tau) (\delta n\delta \rho), \label{eqn:de_dt}
\end{eqnarray}
where $\delta n$ and $\delta\rho$ are, respectively, the differences in the jet production probabilities and the medium densities in the cases of fluctuating and smooth initial conditions. We note that the terms $\rho\delta n$ and $n\delta\rho$ are not present in Eq.(\ref{deviation}) since both vanish after integration over the jet production positions and moving directions. Eq.~(\ref{eqn:de_dt}) indicates that the energy loss difference is determined not only by the fluctuation in the jet production probability density but also by the fluctuation in the local parton density on the jet path \cite{Rodriguez:2010di}.

\begin{figure}[h]
\includegraphics[width=.85\linewidth]{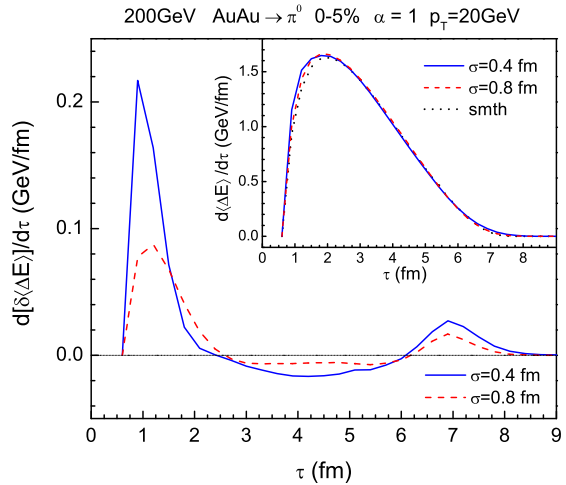}
\caption{\label{fig:en-dt-pt20-05} (Color online). Rate of jet energy loss difference, averaged over all jet paths, in 0-5\% central $Au+Au$ collisions at $\sqrt{s}=200$ GeV for smearing parameter $\sigma=0.4$ fm and 0.8 fm. The inset shows the averaged energy loss rate along the jet path calculated with fluctuating and smooth initial fluctuations.}
\end{figure}

In Fig.~\ref{fig:en-dt-pt20-05}, we show the rate of the energy loss difference for jets with the transverse momentum $p_T$ = 20 GeV, averaged over all jet paths, in 0-5\% central Au+Au collisions at $\sqrt{s}=200$ GeV for smearing parameter $\sigma=0.4$ fm and 0.8 fm. It shows that the correlation between the fluctuation in the production probability of initial parton jets and the fluctuation in the local medium density is positive during the initial stage of jet propagation but changes to negative in the later stage, resulting in enhanced and reduced energy losses, respectively. This result is consistent with that in Ref. \cite{Rodriguez:2010di} for a transversely static medium. The net effect of the fluctuations on the jet energy loss is determined by the sum of the positive and negative differences. As shown in the inset of Fig.~\ref{fig:en-dt-pt20-05}, which gives the averaged energy loss rate along the jet path calculated with fluctuating and smooth initial fluctuations, most energy losses happen close to the initial path of the jet.  Because of the dominance of the initial positive difference, the total energy loss calculated with fluctuating initial conditions is greater than that with smooth initial conditions.

The relation between the jet propagation and the medium evolution can be further clarified if we approximate the time evolution of the parton density $\rho(\tau,r)$ along a jet path in the hydrodynamic evolution of the medium as $\rho(\tau,r) \sim 1/\tau^{\beta}$. According to Eq. (\ref{energyloss}), the time evolution of the medium-dependent jet energy loss can then be simply written as
\begin{equation}
\label{en-alpha-beta}
\Delta E  \sim \tau^{\alpha-\beta},
\end{equation}
if we neglect the small flow effect. As the jet transverses through the medium, its increasing energy loss with the path-length ($\tau^{\alpha}$) is thus suppressed by the decreasing density of the bulk medium
($1/\tau^{\beta}$). Consequently, the total effect of the initial fluctuations on jet quenching is related to the competition between the path-length dependence and the medium-density evolution dependence of the jet energy loss. Since $\alpha$ is always smaller than $\beta$ in our study, the total energy loss mainly takes place during early times when the correlation is positive, thus resulting in more energy loss in the case of fluctuating initial conditions. In non-central collisions, the fireball expands faster than in central collisions ($\beta^{40-50\%} > \beta^{0-5\%}$), so most energy loss happens earlier than in central collisions. As a result, the $R_{AA}$ for the case of initial fluctuations with $\sigma=0.4$ fm decreases by 4\% in central collisions and by 8\% in the 40-50\% centrality of the collisions as shown in Fig. \ref{fig:Raa-rhic-05-4050}.

For the longitudinal expanding medium and the radiative energy loss mechanism considered in Ref.~\cite{Rodriguez:2010di}, $\beta$ is much smaller than $\alpha$, so the total energy loss is dominated by the contribution during later times when the correlation has large negative values. Therefore, the energy loss is smaller and the $R_{AA}$ obtained in Ref.~\cite{Rodriguez:2010di} for the fluctuating initial conditions is larger than in the smooth case.

The path-length dependence of jet energy loss depends on the energy loss mechanism. It is linear ($\alpha$=0) for elastic energy loss, quadratic ($\alpha=1$) for radiative energy loss \cite{Mustafa:2004dr,Adil:2006ei,Qin:2007rn,BDPMS,Zakharov,GLV,GuoW,Wied}, and cubic ($\alpha=2$) for energy loss based on AdS/CFT for the strongly coupled QCD \cite{Dominguez:2008vd,Marquet:2009eq}. How the different power of path-length dependence affects the effect of initial fluctuations on jet energy loss is an interesting question. We illustrate this effect by considering that has either linear or quadratic or other path-length dependence but with the energy loss parameter $\epsilon_0$ in Eq. (\ref{eq:loss}) always determined from the smooth case by fitting the experimental data of the nuclear modification factor in the most central A+A collisions.

\begin{figure}[h]
\includegraphics[width=.89\linewidth]{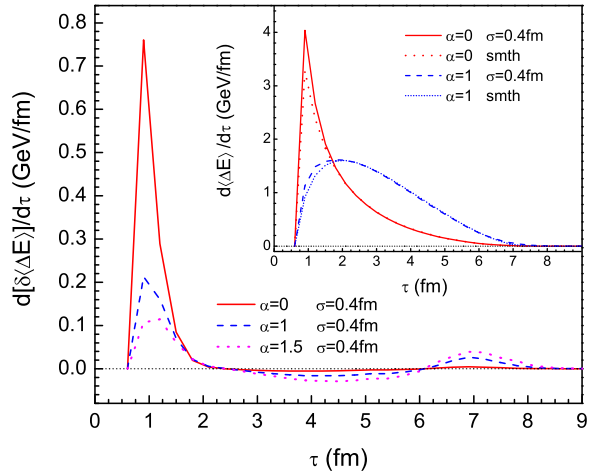}
\caption{\label{fig:en-dt-ww-pt20-tau-alpha} (Color online). Rate of jet energy loss difference averaged over all jet paths in 0-5\% central Au+Au collisions for different powers of the path-length dependence of jet energy loss, $\alpha=$ 0, 1 and 1.5. The inset is the average energy loss rate along the jet path for $\alpha=$ 0 and 1}
\end{figure}

\begin{figure}[h]
\includegraphics[width=.87\linewidth]{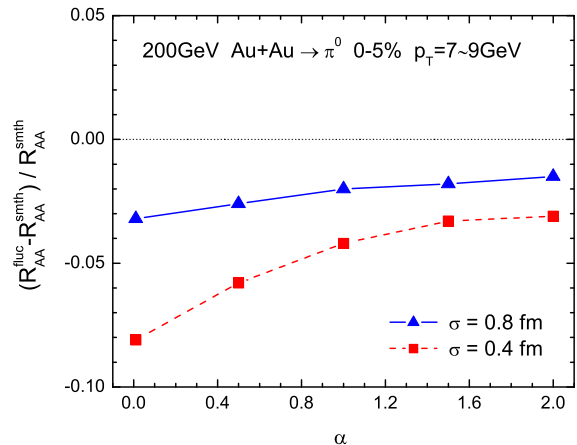}
\caption{\label{fig:Raa-sm.vs.fl} (Color online). Effects of initial fluctuations on jet quenching manifested by the nuclear modification factor of $p_T=7-9$ GeV hadrons as a function of the power index of the path-length dependence of the jet energy loss for three different definitions of the medium density in central Au+Au collisions.}
\end{figure}

In Fig. \ref{fig:en-dt-ww-pt20-tau-alpha}, we show the rate of jet energy loss
averaged over all jet paths in 0-5\% central Au+Au collisions for different powers of the path-length dependence of jet energy loss, $\alpha=$ 0, 1 and 1.5. For the linear path-length dependence ($\alpha$ = 0) of jet energy loss, most energy loss takes place in the initial positive correlation region of the jet path, so the initial positive correlation dominates the fluctuation effect. For the quadratic path length dependence ($\alpha$ = 1), the peak for the jet energy loss rate is shifted closer to the negative correlation region as shown in the inset of Fig.~\ref{fig:en-dt-ww-pt20-tau-alpha}, so the dominance of initial positive correlation is weakened by the negative correlation. Therefore, the energy loss for the linear path-length dependence is greater than the energy loss for the quadratic path-length dependence. This conclusion is supported by the results shown in Fig. \ref{fig:Raa-sm.vs.fl} for the ratio $(R_{AA}^{\rm fluc}-R_{AA}^{\rm smth})/R_{AA}^{\rm smth}$ of $p_T=7-9$ GeV hadrons in central Au+Au collisions as a function of $\alpha$. For the smaller smearing parameter $\sigma$ =0.4 fm, the initial fluctuating conditions decrease the suppression factor by 8\% for the linear path-length dependence of jet energy loss while by 4\% for the quadratic path-length dependence of jet energy loss. The fluctuation effect with the larger smearing parameter $\sigma$ = 0.8 fm is smaller than that with $\sigma$ = 0.4 fm.

\section{Summary}\label{summary}

Based on the 2+1 dimension ideal hydrodynamics, we have studied the effect of initial fluctuations on jet energy loss in relativistic heavy-ion collisions within the description of the NLO
perturbative QCD. Our results show that fluctuating initial conditions lead to slightly more energy loss than smooth initial conditions. In general, the jet energy loss increases with
time due to its path-length dependence but this increase is suppressed by the decreasing medium density with time. Where the total energy loss mainly takes place along the jet path is determined by the competition between the path-length dependence of jet energy loss and the time dependence of the medium density. For fluctuating initial conditions, our results for the rate of the average energy loss difference between the two cases of fluctuating and smooth initial conditions show that the correlation between the fluctuation in the production probability of initial parton jets and the fluctuation in the local medium density is positive during the early times along the jet path and negative during the later times. Consequently, the net effect of initial fluctuations on jet energy loss is determined by whether the energy loss mainly takes place when this correlation is positive or negative. The total energy loss in the fluctuation conditions is then larger than that in the smooth case if most energy loss takes place when the correlation is positive, while it is smaller if it takes place when the correlation is negative. Our results further show that the initial positive correlation dominates the fluctuation effect for linear and quadratic path-length dependence of jet energy loss in central as well as in non-central A+A collisions. However, because this dominance is stronger in non-central collisions than in central collisions, the difference between the nuclear modification factors calculated with fluctuating initial conditions and smooth initial conditions in non-central A+A collisions is greater than that in central A+A collisions. Similarly, the jet energy loss for the linear path-length dependence is more affected by the fluctuation effect than that for the quadratic path-length dependence.

\section*{Acknowledgements}

This work was supported by NSFC of China under Project Nos. 11175071, 10875052 and Key Grant No. 11020101060, and by the U.S. National Science Foundation under Grant No. PHY-1068572, the US Department of Energy under Contract No. DE-FG02-10ER41682, and the Welch Foundation under Grant No. A-1358. H. Z. Zhang thanks members of Cyclotron Institute at Taxes A$\&$M University for their kind hospitality during his visiting stay.


\end{document}